\documentclass{webofc}
\usepackage[varg]{txfonts}   

\wocname{New Frontiers in Physics 2014}
\woctitle{3rd International Conference on New Frontiers in Physics}
\begin{document}
\title{Muon Colliders and Neutrino Factories\\[-.75in]
\rightline{\normalsize\rm IIT-CAPP-14-4}\\[-.05in]
\rightline{\normalsize\rm FERMILAB-CONF-14-516-APC}\\[.25in]}
%
%

\author{Daniel M. Kaplan\inst{1}\fnsep\thanks{\email{kaplan@iit.edu}}\\[0.1cm] for the MAP and MICE Collaborations }

\institute{Illinois Institute of Technology, Chicago, Illinois 60616, USA 
          }

\abstract{%
Muon colliders and neutrino factories are attractive options for future facilities aimed at achieving the highest lepton-antilepton collision energies and precision measurements of Higgs boson and neutrino mixing matrix parameters. The facility performance and cost depend on how well a beam of muons can be cooled. Recent progress in muon cooling design studies and prototype tests nourishes the hope that such facilities could be built starting in the coming decade. The status of the key technologies and their various demonstration experiments is summarized. Prospects ``post-P5'' are also discussed.}
\maketitle
\section{Introduction: Why Muon Colliders are Obviously Best\,---\,and Why Not}
\label{intro}
\begin{figure}[tb]
\centering
\includegraphics[trim=0 10 0 5mm,
width=\linewidth,clip]{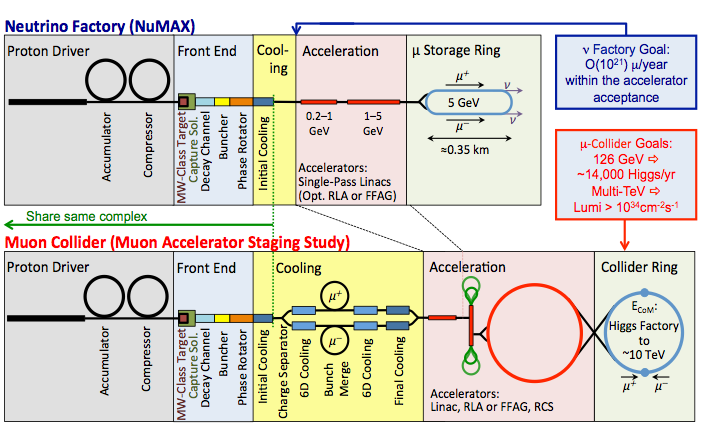}
\caption{Neutrino factory and muon collider conceptual block diagrams. The two types of facilities share a number of elements in common (high-power, medium-energy ``Proton Driver'' and MW-capable target, muon cooling, muon acceleration and storage rings) and, up through the ``Initial Cooling,'' are nearly identical.}
\label{fig:MC}
\end{figure}

Muon colliders offer the only way to study matter with well-understood leptonic probes  both at comparable and at smaller distances than those accessible to the LHC\@. However,
despite the obvious muon advantages at high energies, linear or circular electron--positron colliders are currently under serious consideration to follow up the discovery of the Higgs boson, while muon colliders are not.
Muon advantages stem mainly from its 200-times greater mass:

\medskip
\noindent(1)  Radiative processes (inversely proportional to the fourth power of lepton mass) are greatly suppressed, enabling the use of storage rings and compact recirculating accelerators (Fig.~\ref{fig:MC}). 
\medskip

\noindent(2) So are  the ``beamstrahlung'' interactions that limit 
$e^+e^-$-collider luminosity as energy increases~\cite{Palmer-Gallardo}.  \\[-.1in]

\begin{figure}[htb]
\centering
\includegraphics[width=.55\linewidth,clip]{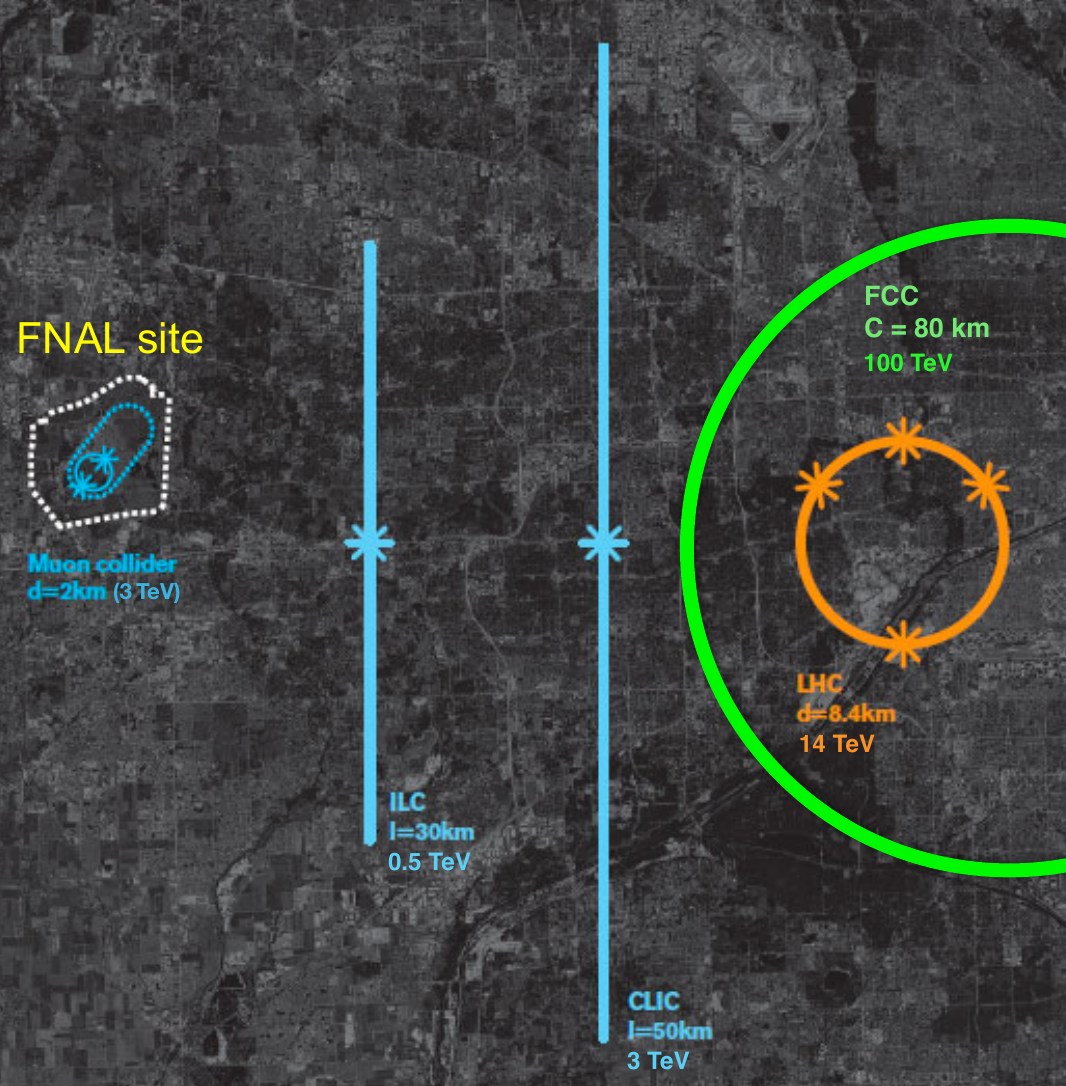}
\caption{Sizes of various proposed colliders compared with FNAL
site. Unlike the others, a muon collider with $\sqrt{s}>3$\,TeV fits on existing
sites.}\label{fig:sizes}
\end{figure}

\noindent (3) The smaller size of a muon collider (Fig.~\ref{fig:sizes}) eases the siting issues and suggests that the cost may be less as well.
\medskip

\begin{figure}[htb]
\vspace{-0.1in}
\centering
\includegraphics[width=.33\linewidth,clip]{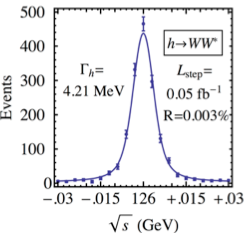}
\includegraphics[width=.45\linewidth]{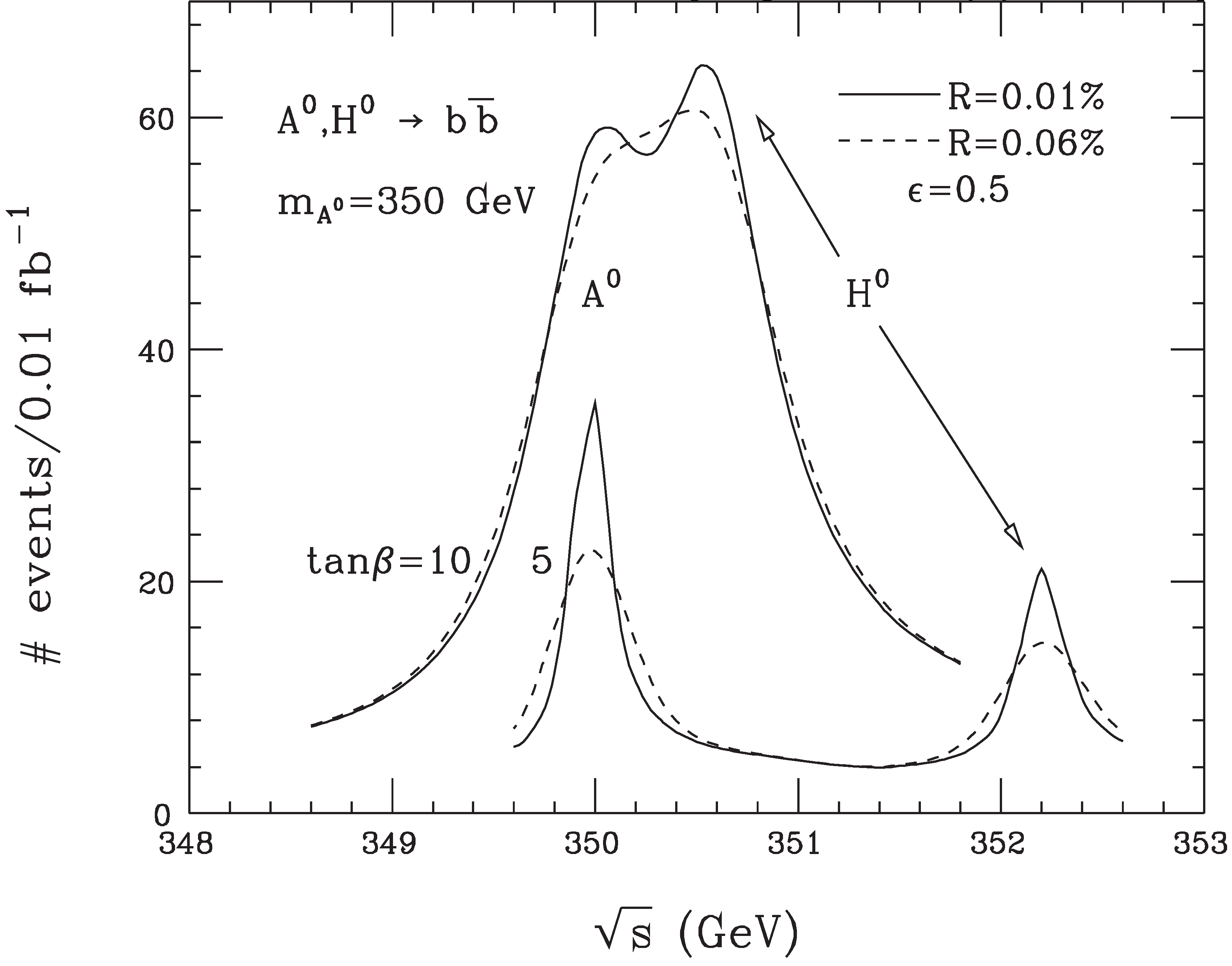}
\vspace{-.05in}
\caption{(left) Example of Higgs resonance scan using $s$-channel production in a muon collider Higgs Factory; (right) resolving scalar and pseudoscalar supersymmetric Higgs partners at a higher-energy muon collider for two possible values of the supersymmetric parameter $\tan{\beta}$~\protect\cite{Barger-Snowmass}. ($R$ is the the collision energy spread.)}\label{fig:Higgs}
\end{figure}

\noindent (4) The cross-section ratio for $s$-channel lepton--antilepton annihilation to scalar bosons, $\sigma_\mu/\sigma_e = ({m_\mu}/{m_e})^2=4.3\times10^4$, gives the muon collider unique access to precision Higgs measurements~\cite{Han-Liu,Bargeretal,Barger-Snowmass,HF}. 
For example, at the $\approx$\,125\,GeV/$c^2$ mass measured by ATLAS and CMS~\cite{LHC-Higgs}, only a muon collider can directly  observe the (4\,MeV) width and lineshape of a Standard Model Higgs boson~\cite{Han-Liu}  (Fig.~\ref{fig:Higgs} left).
\medskip

\noindent (5) Furthermore, should the Higgs have closely spaced partner states at higher mass, only a muon collider has sufficient mass resolution to distinguish them (Fig.~\ref{fig:Higgs} right). (This is a possible feature of supersymmetry as well as other new-physics scenarios.)

\begin{figure}[tbh]
\vspace{-.1in}
\centering
\includegraphics[trim=1 28 511 1mm, width=.49\linewidth,clip]{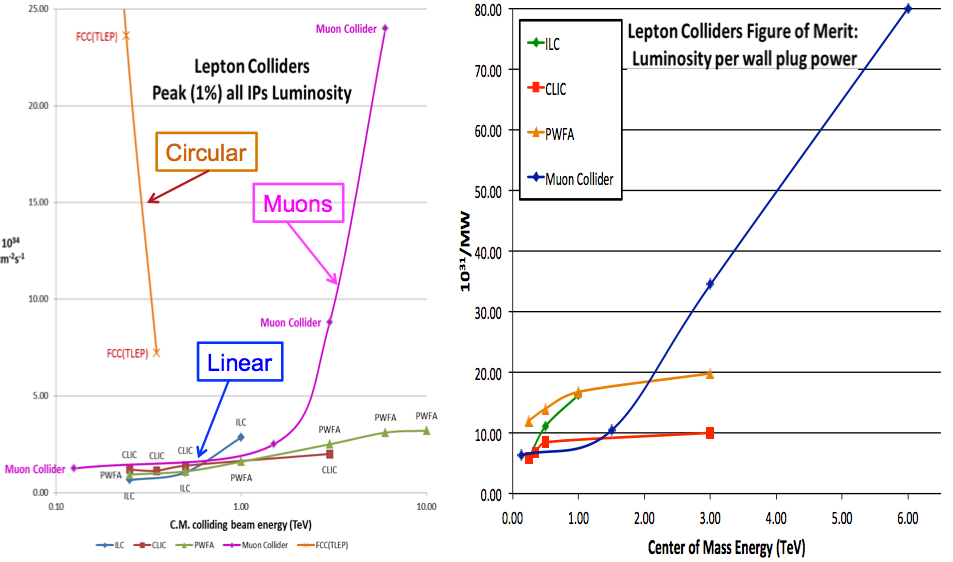}%
\includegraphics[trim=2 22 13 1mm, width=.515\linewidth,clip]{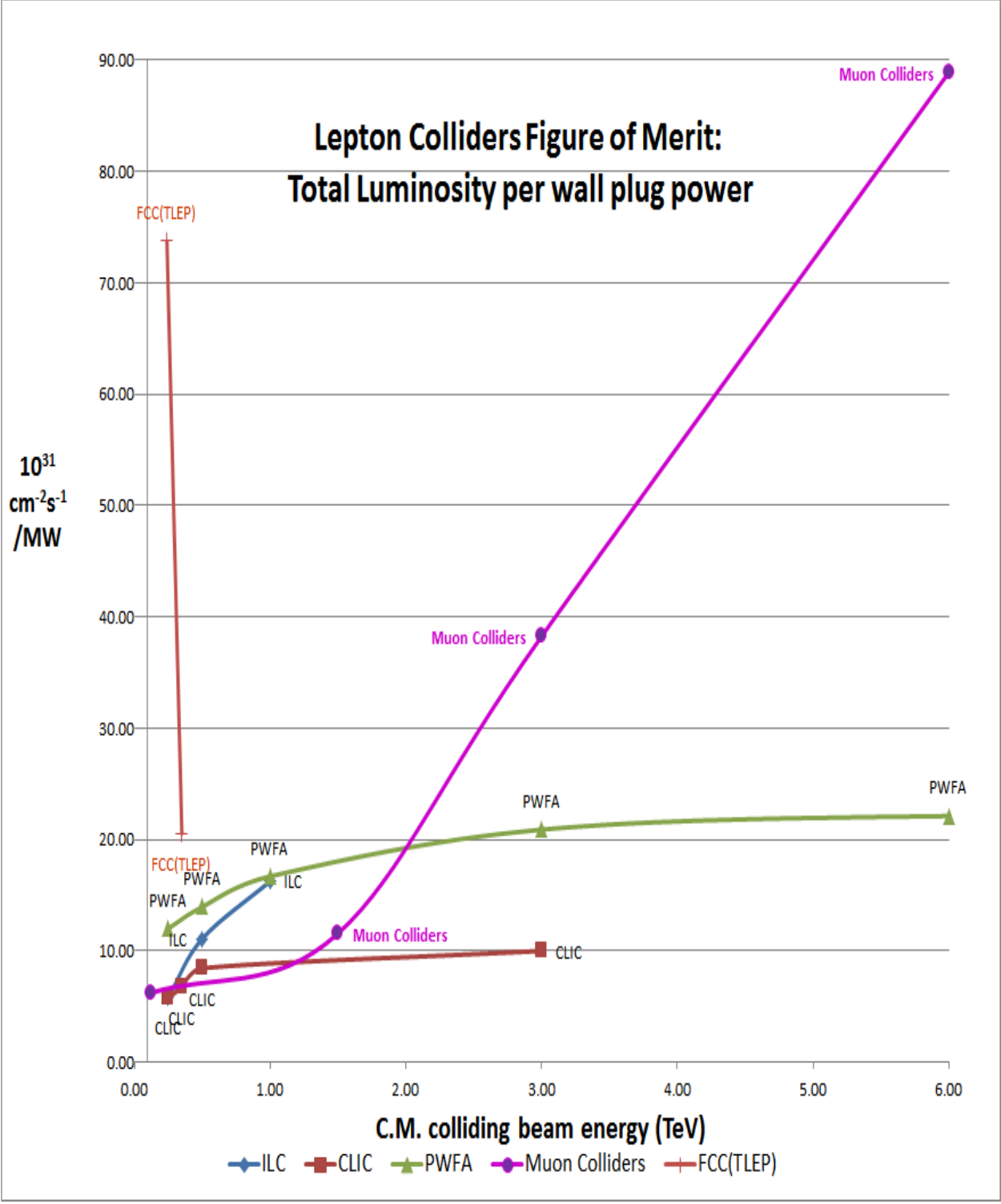}
\vspace{-.1in}
\caption{(left) Luminosity within 1\% of peak $\ell^+\ell^-$ energy, and (right) luminosity per unit wall-plug power, for various proposed collider technologies and lepton choices.}
\label{fig:LC-comparison}
\end{figure}

\newpage
\noindent (6) For the highest lepton energies the muon collider has by far the least operating cost of any proposed approach, and (because of the lack of beamstrahlung) the highest luminosity within 1\% of the energy peak (Fig.~\ref{fig:LC-comparison}).
\medskip

Of course, linear $e^+e^-$ colliders have received enormous attention and resources in recent decades, bringing them to a state of substantial technological readiness. 
And, due to their 2.2\,$\mu$s lifetime, muons must be cooled using novel technology before they can be accelerated to collider energy. This lifetime disadvantage has delayed the general acceptance of muon colliders, inhibiting the R\&D process required to demonstrate that the disadvantage can be overcome. 

\section{Neutrino Factories}

On the other hand, their $\cal{O}(\mu$s) lifetime and simple, well-understood, purely leptonic decay dynamics make muons ideal sources for neutrino beams of unprecedented purity and precision. This realization led to the neutrino factory idea~\cite{Geer}, which has now been brought to the brink of feasibility by the IDS-NF project~\cite{IDS}. On the list of ``go/no-go'' feasibility demonstrations, only the Muon Ionization Cooling Experiment~\cite{MICE} (MICE)  remains to be completed; its progress and prospects are discussed below. Figure~\ref{fig:NF-reach} shows that the neutrino factory has the best precision of any proposed facility for measuring the {\em CP}-asymmetry parameter $\delta$ of the PMNS neutrino mixing matrix, with sensitivity rivaling that in the quark sector\,---\,a reasonable sensitivity goal in order to probe the GUT-scale physics that may link the quark and lepton sectors. Its capability for precision measurement of the PMNS  matrix  gives the neutrino factory the best reach for finding possible new physics beyond three-flavor mixing. It is thus the logical follow-on facility to LBNF.

\section{MASS Facility Staging Plan}
The Muon Accelerator Staging Study~\cite{MASS-White-Paper} (part of the US national Muon Accelerator Program, MAP~\cite{MAP}) has outlined a scenario of neutrino factory and muon collider construction, presumed to be at Fermilab, with successive upgrades, wherein each step is reasonably affordable and brings improved physics reach.

\medskip
\noindent(1) The plan starts with the proposed nuSTORM~\cite{nuSTORM} pion-injected muon storage ring short-baseline experiment, aimed at a definitive test of the sterile-neutrino interpretation of the results from LSND, MiniBooNE,  reactor experiments, etc., as well as precision neutrino cross-section measurements in the energy range crucial to LBNF\@. NuSTORM requires no new technology and no R\&D, so could be built immediately. It will afford the opportunity to develop instrumentation for, and experience with, 
muon storage ring neutrino sources which will be applicable to successor neutrino factories.

\begin{figure}[tb]
\centering
\hspace{-.03in}\includegraphics[trim=0 0 375 0mm,width=.585\linewidth,clip]{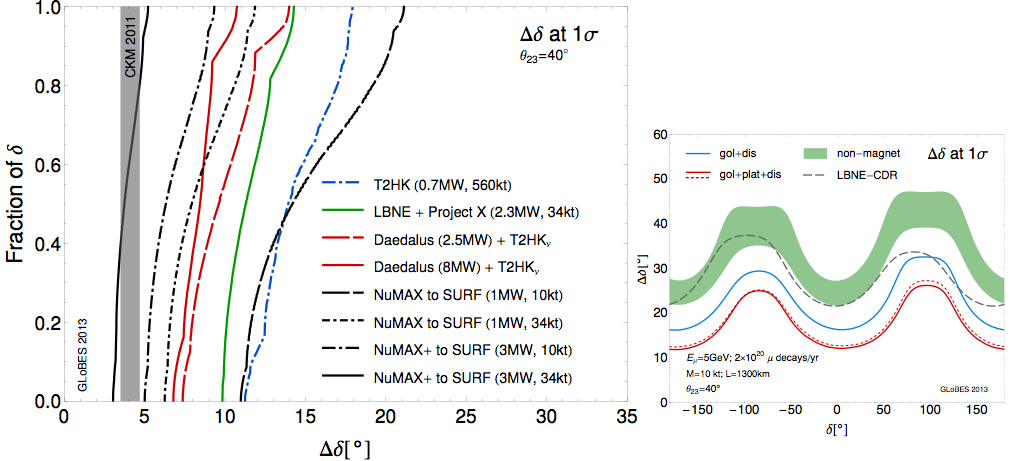}\hspace{-.04in}
\includegraphics[width=.421\linewidth]{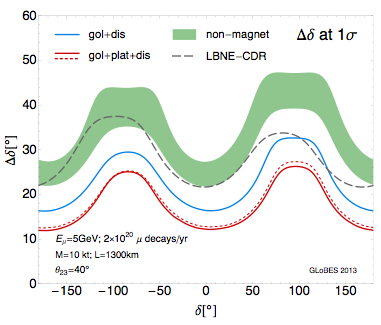}
\caption{(left) Comparison of {\em CP}-violation reach of proposed future neutrino facilities, in terms of the fraction of the range of the {\em CP}-violating phase $\delta$ of the PMNS neutrino mixing matrix over which it can be measured to a given precision, and (right) comparison of  precision achievable in measuring $\delta$ vs.\ its true value. Of all proposed future neutrino facilities, the neutrino factory is seen to have the best {\em CP} reach as well as the best precision on $\delta$.}
\label{fig:NF-reach}
\end{figure}

\medskip
\noindent(2) The next step, NuMAX, is an initial long-baseline neutrino factory at Fermilab optimized for a detector at the Sanford Underground Research Facility (SURF) in South Dakota, with physics reach exceeding that of LBNF (Fig.~\ref{fig:NF-reach}). NuMAX is conceived to start without muon cooling and with a sub-megawatt beam and target.

\medskip
\noindent(3) The follow-on, NuMAX+, facility adds a limited amount of muon cooling and higher-power beam and target, for more than an order-of-magnitude increase in neutrino intensity.
\medskip

Beyond these neutrino-oriented facilities, a series of muon colliders could be built, including:

\medskip
\noindent(4) A Higgs Factory muon collider delivering $>10^4$ Higgs events per year with exquisite energy resolution.

\medskip
\noindent(5) A multi-TeV muon collider (\!$\sqrt{s}\stackrel{<}{_\sim}$\,10\,TeV) offering the best performance and least cost and power consumption of any lepton collider in this energy range (Fig.~\ref{fig:LC-comparison}).
\medskip

 Of course, depending on future physics discoveries and world-wide HEP funding exigencies, the elements of this scheme may not all be built, or not all at Fermilab, and not necessarily in this order. This is also recognized as a rather ambitious plan, extending as it does some 20 to 30 years or more into the future. We consider below how it could be viewed in light of the P5 recommendations~\cite{P5}.
 
The following sections provide a brief overview of these muon facilities as well as some flavor of the R\&D that has been pursued in order to bring them closer to fruition. A more extensive and detailed review may be found in \cite{Palmer-RAST}.

\section{Technical Challenges}
Muon storage-ring facilities present four main technical challenges requiring novel solutions:
(1)~producing enough muons;
(2)~cooling the muon beams to enable high intensity and luminosity;
(3)~rapidly accelerating the beams; and
(4)~storage-ring designs that can deliver small enough $\beta^*$ at the collision points\,---\,or alternatively, the needed neutrino-beam pointing and timing characteristics\,---\,while coping with the high rate of decay electrons.
Solutions have been devised for all four of these challenges. While space constraints prevent a detailed discussion, we here briefly comment on each.

\subsubsection*{Muon production}
\begin{figure}[tb]
\centering
\includegraphics[width=.57\linewidth,trim=0 200 0 200 mm, clip]{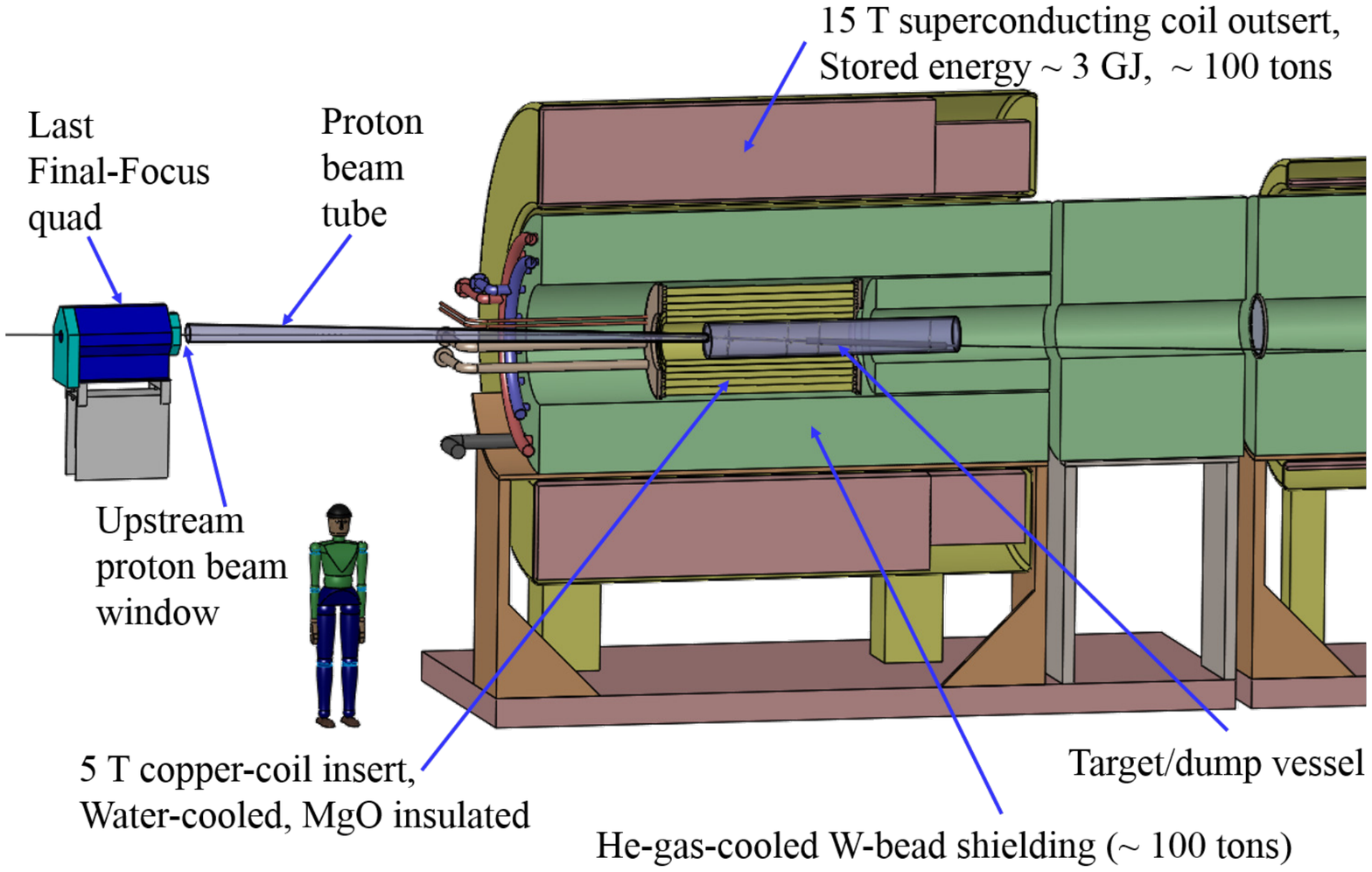}
\includegraphics[width=.42\linewidth,trim=0 0 0 0 mm, clip
]{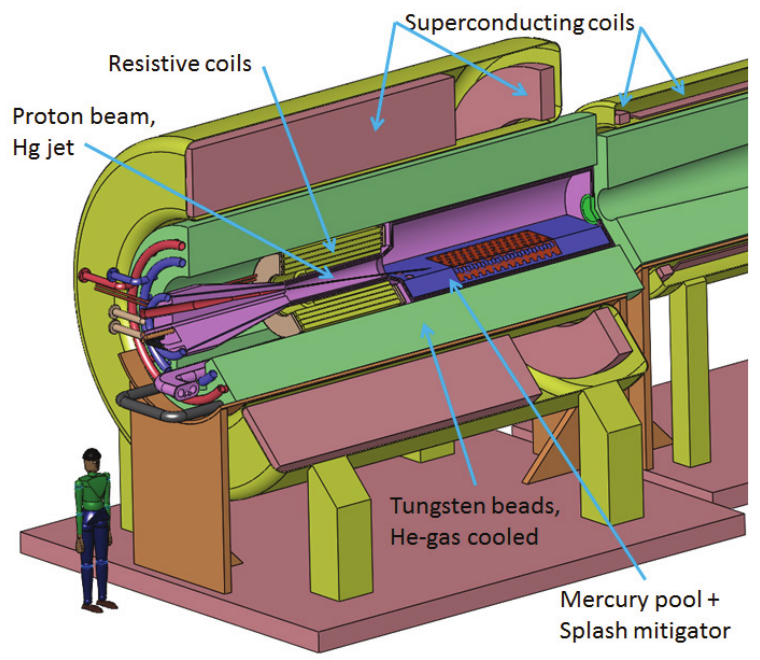}
\caption{Sketch of Target System concept: (left) $\stackrel{<}{_\sim}$\,1\,MW version, with C target~\cite{1MW-IPAC14}; (right) 4\,MW version, with Hg-jet target and Hg-pool beam dump~\cite{Sayed}.}\label{fig:tgt}
\end{figure}

An issue potentially limiting the collider luminosity or neutrino flux that can be achieved is the production of muons in sufficient quantity. The only method that appears suitable is production and decay of low-energy ($<\!1$\,GeV) pions via fixed-target collisions of a megawatt-scale proton beam. Carbon targets have been discussed for beam power up to about 1\,MW~\cite{1MW-IPAC14}. For the one-to-several MW range, a free mercury-jet target has been shown feasible by the MERIT experiment, conducted in 2007 at CERN~\cite{MERIT-IPAC10}. Much subsequent work has gone into optimizing the configuration of the target region and the solenoids (with up to 3\,GJ stored energy) that serve to capture the produced pions and their decay muons~\cite{Ding,Sayed}. Recent concept drawings are shown in Fig.~\ref{fig:tgt}.

\subsubsection*{Muon cooling}
Muon cooling is introduced in some detail in the next section. It is required in order to achieve the luminosity goals for a muon collider, and it is cost-effective for a neutrino factory in that it allows the apertures of the acceleration systems to be reduced. Although earlier neutrino factory work assumed only transverse cooling, recent work has emphasized the utility of muon cooling in all six phase-space dimensions in order to optimize the facility design as a whole by allowing the proton linac to be efficiently reused for the muon beam~\cite{JPD-IPAC2014}.

It is important to distinguish the six-dimensional cooling factor of several million required in order  to achieve high collider luminosity from the much more modest $\sim$\,10 to 50 cooling factor that suffices for a high-intensity neutrino factory. Indeed, a neutrino factory built initially with no muon cooling whatsoever is competitive with proposed future facilities based on neutrinos from pion decay (see Fig.~\ref{fig:NF-reach}). This is one reason why a staged approach such as that discussed above is sensible.

\subsubsection*{Rapid muon acceleration}

To ensure adequate muon survival ($\stackrel{<}{_\sim}$\,10\% decay losses), acceleration must occur at high average gradient. At the low energy (120\,MeV kinetic) that is optimal for ionization cooling, only a linac has sufficient performance. Once the muons have been accelerated to a few GeV, there is sufficient time dilation for recirculating accelerators (with substantially lower costs per GeV) to be used: RLAs, FFAGs, and rapid-cycling synchrotrons, as schematically indicated in Fig.~\ref{fig:MC}. These technologies are challenging to implement for muon applications since, even after cooling, emittances are larger than those in electron and proton machines. At lower energies solenoid focusing is therefore preferred. At higher energy, ``dogbone'' RLAs (with quadrupole focusing) ease switchyard design compared to the more conventional racetracks~\cite{Bogacz-NPB}. The number of passes through each RLA can be increased by means of pulsed quadrupoles~\cite{Bogacz-Max}, and further cost-efficiency can be achieved via two-pass arcs~\cite{Morozov-etal}.

At the highest energies rapid-cycling synchrotrons become favorable. These might employ novel fast-ramping ($\sim$\,kHz) dipoles, with thin, grain-oriented steel laminations~\cite{Summers-Dipole} or ``hybrid'' laminations composed of Si steel with FeCo pole tips~\cite{Witte}. A design has been studied in which such pulsed magnets alternate with fixed-field 8\,T superconducting dipoles, accelerating muons from 30 to 750\,GeV in two rings each the size of the Tevatron\footnote{whose circumference was 6.3\,km.}~\cite{Summers-rings, Berg-Garren}.

\subsubsection*{Storage-ring design}

A neutrino factory requires an oblong (``racetrack'') storage ring, with long straight sections that direct decay neutrinos towards near and far detectors. A series of designs have been developed at various energies, starting in the earliest feasibility studies~\cite{Study-I,Study-II}. Most recently, the IDS-NF study~\cite{IDS} worked out a design for a 10\,GeV decay ring, which can easily be scaled to the 5\,GeV that is optimal for the Fermilab--SURF baseline.

On the other hand, a muon collider storage ring should have minimal straight sections and be as small as possible, in order to maximize the number of turns made by the muons, and hence the number of collisions before they decay. This calls for  high bending field; 10\,T is typical. Designs have been worked through for a 125\,GeV Higgs Factory~\cite{HF-ring} with 4\,MeV energy spread ($\delta E/E\approx 0.003$\%) at the IP and for 1.5~\cite{1.5TeV} and 3.0\,TeV~\cite{3TeV} collision energies.  These employ magnets enclosing tungsten beam-pipe liners in order to absorb decay electrons. At the highest energies care must be taken in order to limit the radiation exposure of people living near locations where the ``neutrino pancake'' due to muon decays in the ring intersects the earth's surface. Thus in the 3\,TeV design, combined-function magnets are used in the arcs instead of quadrupoles in order to have bending field everywhere.

\section{Muon Cooling}
A key ingredient in most of the muon facilities  discussed here is muon cooling\,---\,an area in which there has been important recent progress.
Established (electron, stochastic, and laser) beam-cooling methods take minutes to hours and so are ineffective on the microsecond timescale of the muon lifetime. However,  the muon's penetrating character enables rapid  cooling via ionization energy loss~\cite{ionization-cooling,Neuffer-yellow}. At sufficiently high energy (e.g., a Higgs Factory or higher-energy muon collider), optical stochastic cooling~\cite{Zh-Zo,Nagaitsev-IPAC12} can also be considered and may enable higher luminosity or reduced energy spread. (So-called ``frictional'' cooling has also been considered~\cite{frictional} but appears to be inapplicable to high-intensity stored muon beams and high-luminosity colliders.)

An ionization-cooling channel comprises energy absorbers and radio-frequency (RF) accelerating cavities placed within a suitable focusing magnetic lattice. In the absorbers the muons lose both transverse and longitudinal momentum, and the RF cavities restore the lost longitudinal momentum. In this way, the large initial divergence of the muon beam can be reduced.
Within an energy-absorbing medium, normalized transverse emittance depends on path length $s$ as~\cite{Neuffer-yellow}
\begin{equation} 
\frac{d\epsilon_n}{ds}\approx
-\frac{1}{\beta^2} \left\langle\!\frac{dE_{\mu}}{ds}\!\!\right\rangle\frac{\epsilon_n}{E_{\mu}}
 +
\frac{1}{\beta^3} \frac{\beta_\perp
(0.014)^2}{2E_{\mu}m_{\mu}L_R},\ 
\label{eq:cool} 
\end{equation}  
where $\beta c$ is the muon velocity,  $E_\mu$ the muon energy  in GeV, 
$m_\mu$
its mass in GeV/$c^2$, $\beta_\perp$ the lattice betatron function, and $L_R$ the radiation length of the medium. 
A portion of this cooling effect can be transferred to the longitudinal phase plane (``emittance exchange") by  
placing suitably shaped absorbers  
in dispersive regions of the lattice~\cite{Neuffer-yellow}, by using momentum-dependent path-length within flat absorbers, or within a homogeneous absorber that fills the lattice~\cite{HCC}. 
(Longitudinal ionization
cooling {\em per se}, which would entail operation at momenta above the minimum of the ionization curve, so as to have negative feedback in energy, is impractical due to energy-loss straggling~\cite{Neuffer-yellow}).

\label{sec:matl}
The two terms of Eq.~\ref{eq:cool} 
 represent, respectively, muon cooling by  energy loss and heating by multiple Coulomb scattering. Setting them equal approximates the  equilibrium value of the
emittance, $\epsilon_{n,eq}$, at which the cooling rate reaches zero, and beyond which a given lattice cannot cool. Since the heating term scales with $\beta_\perp/L_R$, to achieve a low $\epsilon_{n,eq}$ requires
low  $\beta_\perp$ at the absorbers. Superconducting solenoids, which can give 
$\beta_\perp<\!\!<1$\,m, are thus the focusing element of choice. Likewise, low-$Z$
absorber media are  
favored, the best being hydrogen (approximately twice as effective for cooling as the next best materials, helium and LiH~\cite{Kaplan-COOL03}). 

\begin{figure}[bt]
\includegraphics[trim=15 80 25 54mm,clip,width=\linewidth]{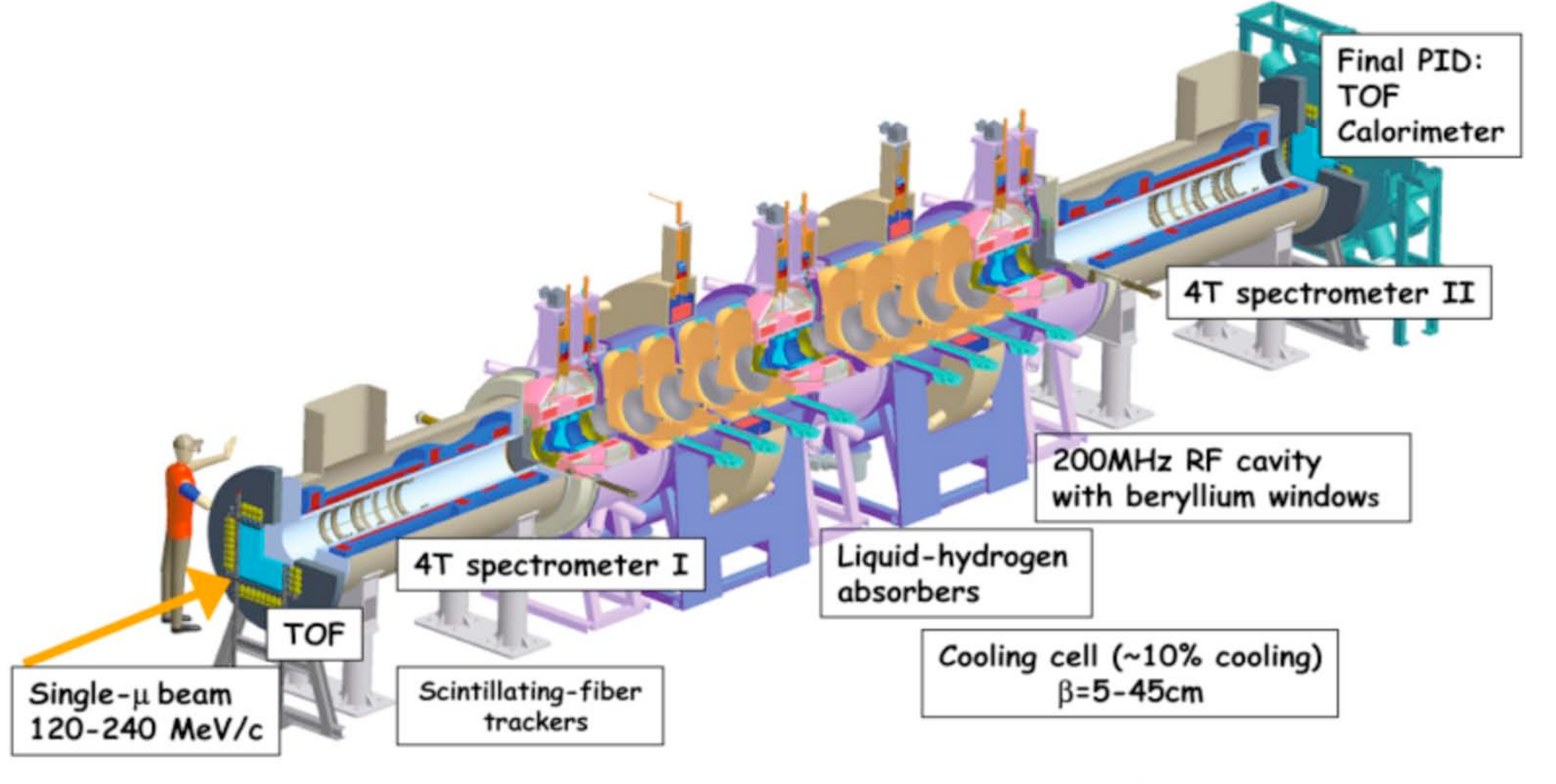}
\caption{Three-dimensional cutaway rendering of MICE apparatus (see text) as originally envisioned: individual muons entering at lower left are measured by time-of-flight (TOF) and Cherenkov counters and a solenoidal tracking spectrometer; then, in cooling section, alternately slowed in LH$_2$ absorbers and reaccelerated by RF cavities, while focused by a lattice of superconducting solenoids; then remeasured by a second solenoidal tracking spectrometer, and their muon identity confirmed by TOF detectors and calorimeters. The cooling section includes three pairs of small ``focus coil" magnets surrounding the absorbers and two large ``coupling coil'' magnets surrounding the RF cavities, comprising one complete lattice cell of the Feasibility Study-II initial cooling lattice, plus one additional absorber and focus-coil pair for symmetry.
}
\label{fig:MICE}
\end{figure}

It is the absorbers that cool the
beam, but for typical ``real-estate'' accelerating gradients ($\approx$\,10\,MeV/m, to be compared with $\langle
dE_\mu/ds\rangle\approx30$ MeV/m for liquid hydrogen~\cite{PDG}), it is the RF cavities that  determine the
length of the cooling channel (see e.g.\ Fig.~\ref{fig:MICE}). 
The achievable RF gradient
thus determines how much cooling is practical before an appreciable fraction of the muons have
decayed. High-gradient vacuum RF cavities (normal-conducting due to the magnetic field in which they must operate) for muon cooling are under development, as is an alternative approach: cavities pressurized with hydrogen gas, thus combining energy absorption and reacceleration~\cite{RF-R&D}. In the first cooling stages the large size of the uncooled beam requires relatively low RF frequency. As the beam is cooled, focal lengths must be shortened in order to reduce the equilibrium emittance, and cavity frequencies and gradients can be increased. Goals are $\stackrel{>}{_\sim}$\,15\,MV/m at 201\,MHz in $\approx$\,2\,T fields, and $\approx$\,25\,MV/m at 805\,MHz in $\approx$\,3\,T\@. Despite early evidence that breakdown limits cavity performance in high magnetic fields, promising results on meeting these goals are now coming from work at the Fermilab MuCool Test Area (MTA)~\cite{MTA}.

In the cooling term of Eq.~\ref{eq:cool}, the fractional decrease in normalized emittance is proportional to the fractional energy loss, thus (at 200\,MeV/$c$) cooling in one transverse dimension by a factor 1/$e$ requires $\sim$\,50\% energy loss and replacement. Ionization cooling thus favors low beam momentum, despite the relativistic increase of muon
lifetime with energy, 
due to the increase of
$dE/ds$ for momenta below the ionization minimum~\cite{PDG}, the greater ease of beam focusing, and the lower accelerating voltage required.  Most muon-cooling designs 
have therefore used momenta in the range 150$-$400\,MeV/$c$. This is also the momentum range in which
the pion-production cross section from thick targets tends to peak and is thus optimal for muon
production as well as  cooling. The cooling channel of Fig.~\ref{fig:MICE}, for example, is optimized for a
mean muon momentum of 200\,MeV$/c$.  

\begin{figure}
\centering
\includegraphics[width=.7\linewidth,trim=0 0 0 3mm,clip]{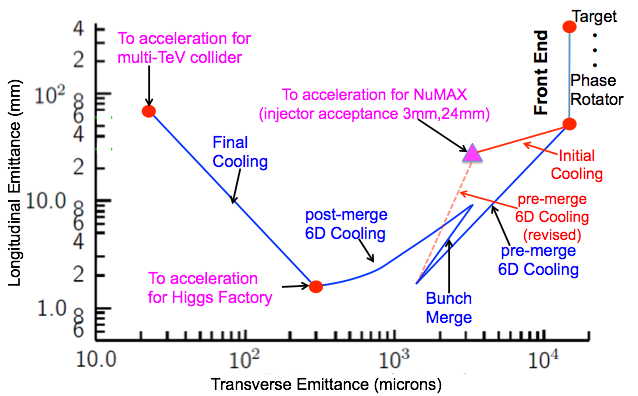}
\caption{Plot of emittance evolution path in longitudinal and transverse planes for representative muon collider cooling scenarios, showing, in blue, the path without a ``dual-use'' linac and, in orange, that with such a linac. In the dual-use linac scenario~\cite{JPD-IPAC2014}, the relativistic part of the Proton Driver H$^-$ linac is reused for medium-energy muon acceleration.}\label{fig:FN}
\end{figure}

\subsubsection*{Muon collider cooling scenarios}
\begin{figure}
\centering
\includegraphics[width=.63\linewidth]{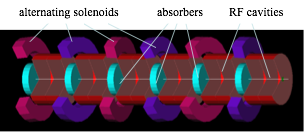}\\ \vspace{0.05in}
\includegraphics[width=.33\linewidth]{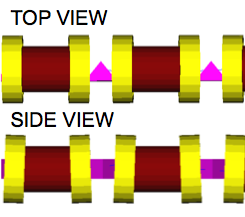}~~
\includegraphics[width=.53\linewidth]{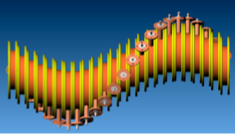}
\caption{Schematic diagrams of (top)  FOFO Snake cooling channel section; (bottom-left) VCC cooling channel section, with superconducting coils in yellow, RF cavities in brown, and wedge absorbers in magenta; (bottom-right)  schematic of HCC cooling channel section, with coils in yellow, cavities in orange, and RF feeds in salmon.}
\label{fig:6D}
\end{figure}

Figure~\ref{fig:FN} shows the emittance evolution in a typical muon collider cooling scenario.  The muon beam emerging from decays of pions produced at the target is captured in solenoids, and bunched and ``phase-rotated'' in order to reduce its energy spread at the expense of increased length~\cite{Phase-rotation}. The bunches then proceed to the initial 6D cooling channel, a candidate for which is the so-called ``FOFO Snake''~\cite{Alexahin-cooling} (Fig.~\ref{fig:6D} top), which is designed to cool both positive and negative muons simultaneously but has limited capability to reach low $\beta_\perp$. Following Initial Cooling the $\mu^+$ and $\mu^-$ bunches will need to be separated for further 6D cooling, then recombined before acceleration and storage; candidate designs to carry out these operations exist~\cite{Separation}.

In Fig.~\ref{fig:FN}, the red point at $\approx$\,1.5\,mm longitudinal emittance is the cooling output point for a Higgs Factory, which needs exquisite energy resolution and, hence, the minimum achievable longitudinal emittance. This is estimated to be limited (in the ``VCC'' design, at least) to $\approx$\,1.5\,mm due to space-charge effects~\cite{Space-charge}. Two cooling approaches (HCC and VCC) have been shown effective in simulation studies aimed at reaching that output point. The VCC (``Vacuum Cooling Channel'') evolved from the ``Guggenheim'' scheme employing helical channels with bending radii large compared to their channel bore, the dipole field components being supplied by tilted thin-lens superconducting solenoid coils~\cite{Gugg}. (The Guggenheim evolved from cooling rings~\cite{Palmer-ring}, which were shown to work but had  injection-kicker issues.) The realization that engineering  such a structure would be challenging led to the current VCC scheme (Fig.~\ref{fig:6D} left), which has very similar simulated performance but in a more straightforward beamline geometry~\cite{Stratakis-etal}.

The competing HCC (``Helical Cooling Channel'') design~\cite{HCC} (Fig.~\ref{fig:6D} right) is a helical structure  with bore  diameter comparable to the bend radius, and is designed to operate with high-pressure gaseous hydrogen distributed throughout. The HCC is believed to work at lower longitudinal emittance than the VCC~\cite{Yonehara}, which might enable a Higgs Factory with even lower energy spread. Additional innovative features of the HCC include RF cavities incorporating dielectric-ring loading for size reduction and helical solenoid magnets composed of current rings that follow the helical paths of the muons~\cite{HCC-eng}. While  pressurized cavities suppress breakdown~\cite{Hanlet}, loading of the cavity by ionization electrons was anticipated to be problematic in pure hydrogen. A dedicated R\&D program at the MTA showed that doping the hydrogen with a percent-level admixture of dry air suffices to suppress this plasma loading, allowing operation at muon collider intensities~\cite{Freeben}.

\subsubsection*{Final Cooling}

For a multi-TeV muon collider, the longitudinal emittance at the Higgs Factory cooling output point is much smaller than necessary, while the transverse emittance is too large for the desired ${\cal O}(10^{34})$ luminosity. This emittance mismatch is alleviated via ``Final Cooling,'' in which the muon energy is allowed to fall in order to take advantage of the rising $dE/dx$ curve at low energy, and the cooling-channel equilibrium emittance is further reduced by means of small-bore 30--40\,T solenoids enclosing LH$_2$ absorbers~\cite{Sayed-FC}. Such magnets appear to be feasible and are being developed by NHMFL~\cite{NHMFL} among others~\cite{BNL}. Alternatives to Final Cooling have also been discussed, incorporating, e.g., ``Parametric Ionization Cooling''~\cite{Maloney} or a ``potato slicer'' emittance exchanger~\cite{potato}.

\section{MICE}
The Muon Ionization Cooling Experiment, after  delays associated with building large superconducting magnets to be cooled by closed-cycle cryocoolers, is on track to take first measurements with absorbers in the beam in 2015. One ``lesson learned'' (which was already obvious to the experts some years ago) is to use large helium refrigerators in any real cooling channel\,---\,although this option was unavailable to us at Rutherford Appleton Laboratory (RAL), where MICE is sited. A second is to move to higher RF frequency so as to reduce the transverse size of components. Simulation studies have now shown that 325\,MHz RF cavities (rather than the 201\,MHz ones used in MICE) have sufficient aperture, even at the large $\cal O$(10$\pi$)\,mm$\cdot$rad RMS normalized transverse emittance of an early-stage muon cooling lattice. A third lesson is to avoid whenever possible large (``coupling coil'') superconducting magnets surrounding the cavities (see Fig.~\ref{fig:MICE}), and cooling lattices without such coils have now been developed and shown to deliver good performance~\cite{Stratakis-etal}.

\begin{figure}
\centering
\includegraphics[width=\linewidth]{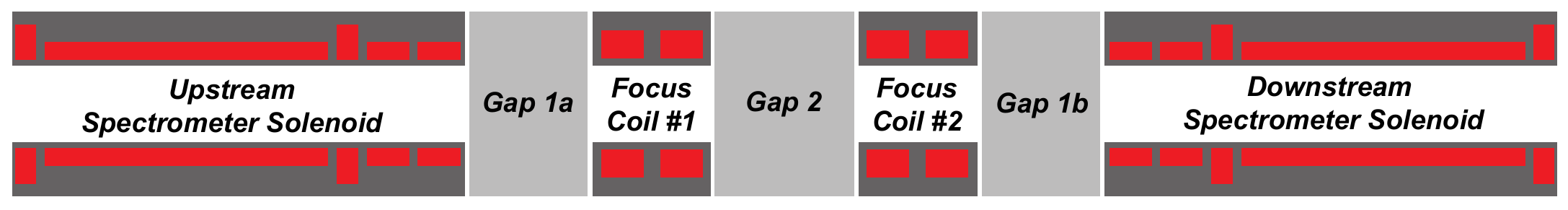}
\caption{Generic schematic for MICE Final Cooling Demonstration, containing three gaps into which RF cavities and/or absorbers may be placed.}\label{fig:pi}
\end{figure}

The principle of MICE has been to develop very precise emittance measurement techniques, with a low enough beam intensity that each muon can be tracked individually, so as to avoid the need for a long and expensive cooling  section. Thus MICE as originally proposed~\cite{MICE} (Fig.~\ref{fig:MICE}) included  just one lattice cell of the 201\,MHz lattice from Neutrino Factory Feasibility Study-II~\cite{Study-II}. As of this (Oct., 2014) writing, with the US ``P5'' committee having recommended an early termination of MICE~\cite{P5}, a new and simpler lattice is being devised in order to obviate the need for the coupling coils. A generic diagram of the new arrangement, shown in Fig.~\ref{fig:pi}, is more reminiscent of recent lattice designs, such as that of the IDS-NF~\cite{IDS}, than of the Study II design. Preliminary simulation studies indicate a transverse cooling factor on the order of several percent, easily measurable in MICE given the 0.1\% emittance resolution provided by the scintillating-fiber tracking systems~\cite{Tracking} of the input and output solenoidal spectrometers. The MICE ionization cooling demonstration using the arrangement of Fig.~\ref{fig:pi} is now scheduled for data-taking by 2017.

\section{Conclusions and Perspective}

The muon collider/neutrino factory intellectual journey has been an exciting and a fascinating one, starting from the earliest suggestions~\cite{Tikhonin-Budker}, accruing important innovations~\cite{MC}, and culminating in the sophisticated simulation studies of today~\cite{Stratakis-etal}. Indeed, we have reached a point at which the muon collider and neutrino factory concepts,  {\em a priori} seemingly unlikely, now increasingly appear  feasible. 
Moreover, the neutrino factory has been shown to be the most powerful way to study the only non-Standard Model physics that is definitively established, neutrino oscillation. 

Should LHC discover a new scale of phenomena above 1\,TeV, a muon collider will be the obvious way to study it with precision.\footnote{$e^+e^-$ colliders being  less suited to such a mass range: CLIC, aimed at this mass range, has been shown to be immensely complex and expensive both to build and to operate, as well as suffering from beamstrahlung as discussed above.} Absent such discovery, construction of a large-scale stored-muon facility may be farther off in the future, with a neutrino factory the likely follow-on (some two decades hence) to LBNF\@. The work briefly summarized and cited here will have paved the way to these  future machines. Smaller-scale implementations of cooled muon beams have also been discussed~\cite{Yoshikawa} and might proceed on other grounds.

Following nearly two decades of inspired work by the Muon Collaboration, the Neutrino Factory and Muon Collider Collaboration~\cite{NFMCC},\footnote{As the Muon Collaboration renamed itself upon realization of the power of the neutrino factory concept.} Muons, Inc.~\cite{Muonsinc}, the Muon Collider Task Force~\cite{MCTF}, and now the Muon Accelerator Program, the P5 committee has recommended the termination of this effort\,---\,albeit, with possible  support for ongoing concept (though not technology) development through the DOE's General Accelerator R\&D program. The quest to do more with muons remains close to the hearts of its devotees, and physics soon to be discovered may yet have the last word.

\section*{Acknowledgments}
It is a pleasure to acknowledge the many MC, NFMCC, Muons, Inc., MCTF, MICE, and MAP collaborators, too numerous to name here, without whom the work described could not have taken place, as well as the conference organizers who so kindly invited me to participate in ICNFP2014. This work is supported by the US  Dept.\ of Energy (via the Muon Accelerator Program) and by the National Science Foundation under award PHY-1314008.
%
%
%

\end{document}